\documentclass[a4paper,11pt]{article}
\pdfoutput=1 

\usepackage{jinstpub} 

\usepackage{DRtubes-defs}
\usepackage[T1]{fontenc} 
\usepackage{xspace}
\usepackage{xcolor,colortbl}
\usepackage{hyperref}
\usepackage{subfigure}
\usepackage{comment}
\usepackage{tikz}
\usepackage{doi}
\usepackage{lineno}

\title{Energy Response and Resolution to Positrons in a Capillary-Tube Dual-Readout Calorimeter}

\author[a]{S.F. Albergo}
\author[b]{A. Braghieri}
\author[c,d]{A. Burdyko}
\author[e,f]{Y. Cai}
\author[c,g]{L. Carminati}
\author[g]{E. Delfrate}
\author[e]{D. Falchieri}
\author[b]{R. Ferrari}
\author[b]{G. Gaudio}
\author[e]{P. Giacomelli}
\author[h]{A. Loeschcke Centeno}
\author[c,g]{E. Mazzeo}
\author[a]{S. Millesoli}
\author[c,g]{L. Nasella}
\author[b,i]{A. Negri}
\author[b,i]{A. Pareti}
\author[a]{R. Persiani}
\author[e]{L. Pezzotti}
\author[b]{G. Polesello}
\author[h]{P.F. Salvatore}
\author[c,d]{R. Santoro}
\author[b,i]{L.D. Tacchini}
\author[c,g]{R. Turra}
\author[b]{N. Valle}
\author[e,f]{I. Vivarelli}

\affiliation[a]{INFN, Sez. Catania}
\affiliation[b]{INFN, Sez. Pavia}
\affiliation[c]{INFN, Sez. Milano}
\affiliation[d]{Universit\`a dell'Insubria, Como}
\affiliation[e]{INFN Sez. Bologna}
\affiliation[f]{Universit\`a Alma Mater Studiorum, Bologna}
\affiliation[g]{Universit\`a Statale, Milano}
\affiliation[h]{University of Sussex}
\affiliation[i]{Universit\`a degli Studi di Pavia}

\emailAdd{iacopo.vivarelli@unibo.it}

\abstract{We present the results of a test beam campaign on a capillary-tube fibre-based dual-readout calorimeter, designed for precise hadronic and electromagnetic energy measurements in future collider experiments. The calorimeter prototype consists of nine modules, each composed of brass capillary tubes housing scintillating and Cherenkov optical fibres, read out using silicon photomultipliers for the central module and photomultiplier tubes for the outer modules. The performance of the detector was assessed using a positron beam with energies ranging from 10 to 120 GeV at the CERN SPS H8 beamline. The prototype is characterised in terms of the linearity and resolution of its energy response to positrons. The results confirm the feasibility of the capillary-tube mechanical design for large-scale dual-readout calorimetry and provide a benchmark for future detector development within the HiDRa project.}


\begin{document}

 \maketitle

\flushbottom

\section{Introduction}
\label{sec:introduction}
Dual-readout calorimetry~\cite{Lee:2017xss} is a compelling technique, actively investigated by several groups as an option to reach excellent hadronic calorimetric energy resolution at future lepton colliders, (such as, for instance, \mbox{FCC-ee} and CEPC \cite{FCC:2018evy,CEPCStudyGroup:2018ghi}). 

Dual-readout calorimetry operates on the principle of dual signal sampling utilising two distinct sensitive materials characterised by differing $h/e$ ratios. By combining the information from these two signals, this approach effectively compensates for fluctuations in the electromagnetic fraction of hadronic showers, thereby significantly enhancing the resolution of energy measurements and restoring linearity in the calorimeter response to hadrons. This technique is well-established~\cite{Lee:2017xss}, with its feasibility confirmed through an extensive experimental programme spanning two decades~\cite{Akchurin:2005eu,Akchurin:2005an,Akchurin:2005rs,Akchurin:2013yaa,Lee:2017shn, Akchurin:2014aoa}. The outcome of this programme is a design that incorporates two types of optical fibres embedded within an absorber, oriented nearly parallel to the trajectory of incoming particles. Scintillating fibres sample the charged particles in the shower, while undoped plastic fibres collect Cherenkov light predominantly produced by electrons and positrons, providing sensitivity to the electromagnetic shower component.

Recent advancements in dual-readout technology include the integration of silicon PhotoMultipliers (SiPMs) as light detectors capable of reading individual fibres~\cite{Antonello:2018sna}. Simulations of a comprehensive 4$\pi$ dual-readout calorimeter\footnote{In this case, the mechanical and geometrical configurations were different from the one with capillary tubes discussed in this paper. Still, the results are worth to be mentioned here as a benchmark of what can be achieved with a dual-readout calorimeter. A full simulation based on the key4hep framework, is now available~\cite{IDEAStudyGroup:2025gbt}, and these studies will be repeated in the future.} have been documented both using the calorimeter as a standalone device to measure the energy of both electrons/photons and hadrons \cite{lorenzoPhD, Pezzotti:2022ndj}, and in conjunction with a crystal-based dual-readout electromagnetic section positioned ahead of the fibre-based hadronic calorimeter~\cite{Lucchini:2020bac, Lucchini:2022vss}. The combined crystal + fibre-based calorimeter shows great potential in terms of energy resolution to hadrons: Ref.~\cite{Lucchini:2022vss} estimated the performance of a simple particle flow algorithm on top of the dual-readout calorimeter response, obtaining jet energy resolutions of $\sigma/E \sim 30\%/\sqrt{E\ \mathrm{[GeV]}}$. The combined crystal + fibre  configuration is now the baseline for the IDEA detector concept~\cite{IDEAStudyGroup:2025gbt}. 

This study focuses on a test-beam done in 2023 of a prototype designed with a recently explored mechanical construction concept: optical fibres are housed in individual cylindrical brass capillary tubes, which are then glued together to form calorimeter modules. This design offers a cost-effective and flexible solution for large-scale construction. The prototype construction is documented in Ref.~\cite{Karadzhinova-Ferrer:2022paf}. It represents a milestone in terms of development of the mechanical construction technique, and in terms of the granularity of the readout. Its size guarantees a good  containment for electromagnetic showers (94\% for electrons with an energy of 20 GeV), but only a poor one for hadronic showers. A first assessment of the quality of the prototype response to positrons was performed with test-beam data in 2021, documented in Ref.~\cite{Ampilogov:2023zxb} in terms of linearity and resolution of the energy measurement, and of the quality of the shower profile measurement. However, a poor positron beam purity and a non-optimal placement of one of the auxiliary detectors limited the ability to assess the energy resolution at positron beam energies up to $\Ebeam  = 30\ \textrm{GeV}$. Moreover, the lack of a vertical tilt angle between the beam and the calorimeter axis introduced a dependency of the response on the particle impact point on the calorimeter, which had to be corrected at the analysis level. These issues forced the use of the software simulation to make a statement on the optimal calorimeter electromagnetic energy resolution. 

All these issues were solved in a second test-beam, performed in 2023 at the H8 beam line at the CERN SPS. This paper describes the results of this second test-beam, in terms of linearity and resolution of the prototype energy response to positrons from 10 to 120 GeV.  

Section~\ref{sec:det_descr} provides an overview of the experimental setup, including the calorimeter structure, readout system, and auxiliary detectors used to efficiently select  positron interactions. The optimisation of the positron selection and estimated positron beam purities are described in Section~\ref{sec:part_selection}. The calibration procedure, detailing the equalisation of module responses and calibration of the electromagnetic scale, is outlined in Section~\ref{sec:calibration}. The calorimeter response to positrons is presented in Section~\ref{sec:results}, with concluding remarks in Section~\ref{sec:conclusions}.

\section{Experimental setup}
\label{sec:det_descr}
The prototype tested on beam is the same as the one tested in 2021, and described  in Ref.~\cite{Ampilogov:2023zxb}. 

\begin{figure}[htb]  
\begin{center}
\includegraphics[width=0.85\textwidth]{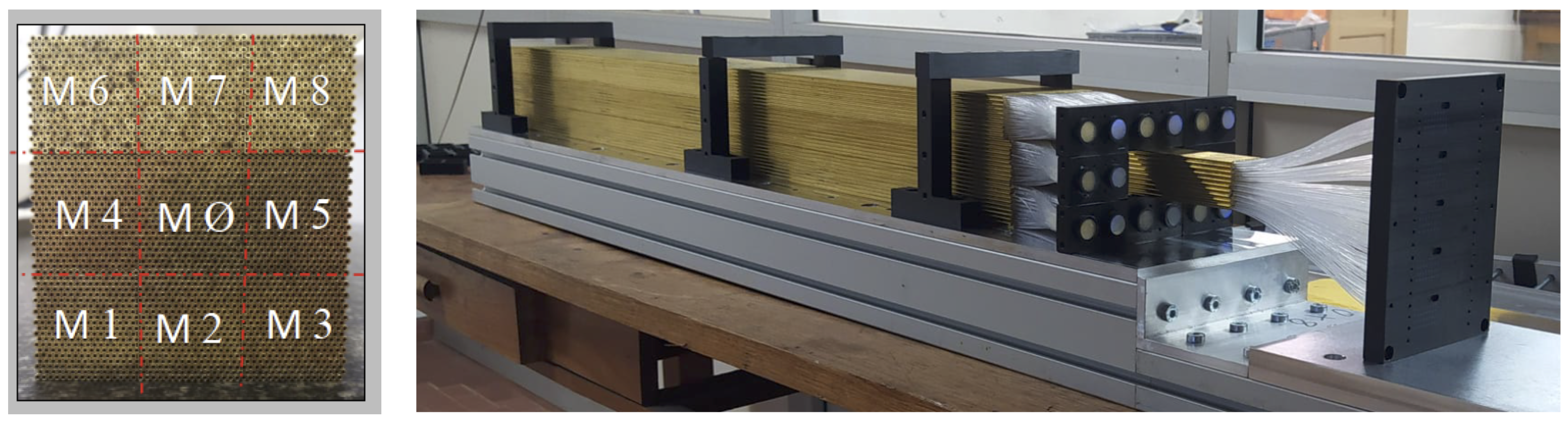}
\end{center}
\caption{View of the prototype and its segmentation in the $\M{0}-\M{8}$ modules.}
\label{fig:prototype}
\end{figure}  

Nine identical modules, labelled as $\M{0}-\M{8}$, are arranged as shown in Figure~\ref{fig:prototype}. Each module is 100-cm long and its dimensions transverse to the beam are $3.3\times 3.3\ \mathrm{cm^2}$, yielding a total prototype size of about $10 \times 10 \times 100\ \mathrm{cm}^3$. Each module is assembled by gluing together 320 100-cm long brass (63\% Cu, 37\% Zn) capillary tubes. Each tube encloses an optical fibre. The dual readout is obtained by utilising two different sets of fibres: one set of scintillating and one of clear undoped fibres. The scintillating fibres (BCF-10 from Saint Gobain~\cite{6551127}, now Luxium) have a polystyrene-based core and a single PMMA clad. The emission peak is at 432 nm, and the light yield is about 8000 photons per MeV. The clear undoped fibres (referred to as ``Cherenkov'' in the following) are SK-40 from Mitsubishi~\cite{SK40}, with  a PMMA resin core and a fluorinated polymer clad.

\begin{figure}[htb]  
\begin{center}
\includegraphics[width=0.45\textwidth]{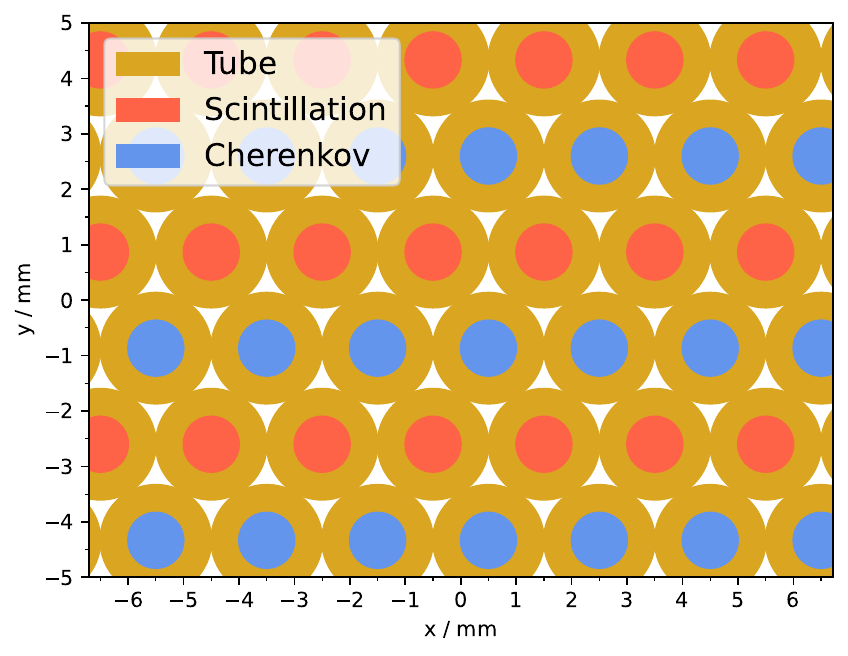}
\end{center}
\caption{Sketch of the front face of the calorimeter detailing the relative positions of the Cherenkov and scintillating fibres.}
\label{fig:prototype_sketch}
\end{figure}

Overall, the volumes of the prototype are 66\% brass, 22\% fibres, with air and glue covering the rest. The effective radiation length is estimated to be 22.7 mm, while the Moli\`ere radius is 23.8 mm. The alternating layout of the scintillating and Cherenkov fibres is shown in Figure~\ref{fig:prototype_sketch}. 

The external modules $\M{1}-\M{8}$ are instrumented with Hamamatsu PMTs of the R8900 series~\cite{R8900}. The scintillating and clear fibres are separated and bundled in two groups on the back side of each module to match the PMTs' window. A yellow filter (Kodak Wratten 3, with nominal transmission of about 7\% at 425~nm and 90\% at 550~nm) is placed between the scintillating fibres and the PMTs to attenuate the scintillation signal and to cut off short wavelength components of the light: although not necessarily relevant for positrons, this helps reducing the calorimeter response dependence on the shower depth and starting point by selecting wavelengths with a longer fibre attenuation length. The PMTs are read out with V792AC QDC modules produced by CAEN S.p.A..

Each individual fibre of the central module $\M{0}$ is instead read out by an individual SiPM with a $1.3 \times 1.3\ \mathrm{mm^2}$ sensitive area. The SiPMs (S14160-1315 PS~\cite{S14160}) have a pitch of $15\ \mathrm{\mu m}$, for a total number of cells of 7284.
The fibres at the back of the calorimeter drive the light to front-end boards, each hosting 64 SiPMs. The front-end board is split in two optically-insulated groups to avoid optical cross-talk between the Cherenkov and scintillation light. As for the modules $\M{1}-\M{8}$, yellow filters are placed between the scintillating fibres and the SiPMs. In addition, for $\M{0}$ optical grease (Saint Gobain BC-630  Silicone Optical Grease - 95\% transmission between 280 and 700 nm) is used to improve the optical transmission between the fibres and the SiPMs. 

The SiPM readout is based on the Front-End Readout System (FERS) produced by CAEN S.p.A.~\cite{FERS:CAEN}. Each readout board (A5202) is equipped with two Citiroc~1A~\cite{CITIROC} to operate 64 SiPMs. Five FERS in total are used to read out the 320 SiPMs. The signal produced by each SiPM feeds two~charge~amplifiers (named High Gain, or HG, and Low Gain, LG in the following) with tunable gains. The gain of the HG was set to be roughly 25 times that of the LG. Both signals are read out simultaneously and stored on disk. The settings for the two charge amplifiers were chosen to guarantee the ability to have a good quality measurement of the multiphoton spectrum in HG and a wide dynamic range, while maintaining an overlap between the signals acquired with the two different gains to be used for their mutual calibration. The settings chosen allowed signals from 1 to almost 4000 photoelectrons (\phe) to be read out. This corresponds to  about 55\% of the SiPM occupancy considering the microcells available in the sensitive area. The FERS system reading the SiPM from $\M{0}$ had an internal self-trigger system which was activated when more than 3 SiPMs (over the 64 served by one A5202 board) exceed the discriminator threshold set at 2.5 \phe. The data were read in presence of a coincidence between the internal and physics (see later) triggers. This caused no bias on the energy readout, but required a dedicated approach to study the SiPM noise contribution, as discussed in Section~\ref{sec:pedestals}. 

\subsection{Beam setup}

A set of auxiliary detectors present on the beam line was used as trigger system and to help particle identification. The setup was similar to that described in Ref.~\cite{Ampilogov:2023zxb}. Figure~\ref{fig:beam_sketch} sketches the beam setup. 

\begin{itemize} 

\item Upstream of the beam, two Cherenkov threshold counters~\cite{Dannheim:2013iea} were available. The pressure of the $\mathrm{He}$ gas was set to optimise the separation between  electrons and pions depending on the beam energy. 

\item A system of three scintillators was used to trigger on beam particles. The coincidence of $\mathrm{T_1}$ and $\mathrm{T_2}$, each 2.5~mm thick, with an area of overlap of about  $4 \times 4\ \mathrm{cm^2}$ was used in anti-coincidence with a third scintillator counter ($\mathrm{T_3}$, also 2.5~mm 
 thick), placed downstream the beam. 
$\mathrm{T_3}$ had a 10-mm radius hole in its centre: its purpose was to veto off-axis particles. Therefore, the combination $\left(\mathrm{T_1} \land \mathrm{T_2}\right) \land \bar{T}_3$ defined what will be referred to as the ``physics trigger'' in the following. 

\item A pair of Delay Wire Chambers (DWC1 and DWC2) were placed upstream and downstream the beam with respect to the trigger scintillators. They were used to determine the location of the impact point of the particles at the calorimeter front face. The typical precision that could be achieved was of a few mm. 

\item A preshower detector (PS in the following), consisting of 5~mm of lead and a scintillator slab read out with a photomultiplier, was located at 10~cm from the face of the calorimeter. A high-purity electron/positron selection can be achieved by requiring a signal higher than that of a Minimum Ionising Particle  (MIP) in the scintillator. 


\begin{figure}[htb]
\begin{center}
\includegraphics[width=1.05\textwidth]{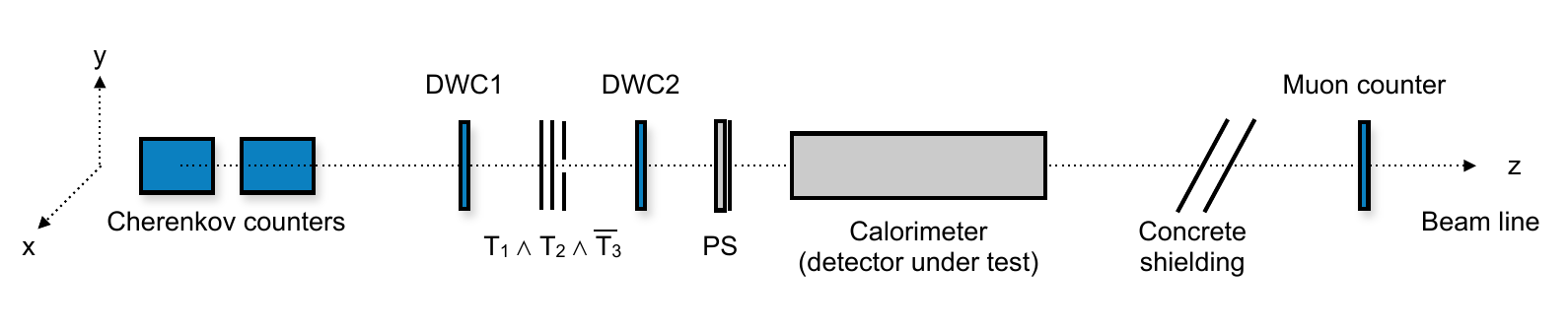}
     \end{center}
\caption{Sketch of the beam line setup. The diagram is not to scale.}
\label{fig:beam_sketch}
\end{figure}  

\item About $20\ \mathrm{m}$ downstream of the calorimeter, behind the concrete shielding, a $50\times 50\ \mathrm{cm^2}$ scintillation counter $\mathrm{T_{\mu}}$ served to identify the muons in the beam. 

\end{itemize}

Every ten physics triggers, a random ``pedestal'' trigger was produced. All trigger signals, physics and pedestal, were sent to two data acquisition systems, one reading the auxiliary detectors and the PMTs of modules $\M{1}-\M{8}$, and one reading the signals from the SiPMs of the $\M{0}$ module. 

The synchronisation of the two data acquisition systems was done offline, by making use of the pedestal events.  

A right-handed orthogonal system of coordinates with the $z$-axis along the beam line, and with the $y$-axis pointing upwards is used in the remainder of this paper. The origin of the coordinate system  is on the front face of the calorimeter, at the geometrical centre of $\M{0}$. The calorimeter prototype was placed on beam so that its longest side formed an angle of about $2^{\circ}$ with the $z$-axis in both the $x-z$ and the $y-z$ plane. This is to avoid channeling effects (particles entering and travelling long distances in an optical fibre) and to minimise any dependence of the calorimeter response on the impact point (discussed extensively in Ref.~\cite{Ampilogov:2023zxb}).

\section{Particle selection}
\label{sec:part_selection}
A pure beam of positrons is needed  for the calibration of the prototype and for the assessment of its performance.
This requires the identification and removal of interactions due to particles other  than positrons.

The positron selection starts from the DWC detectors, which are used to avoid selecting particles that hit the calorimeter front face too far off its centre, leading to additional lateral leakage. The calibration and alignment of the DWCs to the calorimeter prototype is done by exploiting the excellent lateral granularity of the calorimeter: the position $\left(X_{\mathrm{calo}},Y_{\mathrm{calo}}\right)$ of the shower barycentre is evaluated as 

\begin{align*}
    \left(X_{\mathrm{calo}},Y_{\mathrm{calo}}\right) = \left(\frac{\sum_i E_i x_i}{\sum_i E_i} , \frac{\sum_i E_i y_i}{\sum_i E_i}\right)
\end{align*}

\noindent where $x_i$ and $y_i$ are the horizontal and vertical coordinates of the $\M{0}$ $i$-th fibre with respect to the tower centre, and $E_i$ is the energy deposited in the fibre. The prototype is positioned so that the centre of the beamspot (as measured in the prototype)  coincides with the $\M{0}$ geometrical centre. The alignment with the DWCs is done by imposing that the centre of the beamspot as determined from the DWCs coincides with that determined from the calorimeter.   

A circle with radius 6.5 mm in the beamspot centre is selected in each DWC. Particles travelling at an angle with respect to the beam line are suppressed by requiring that the coordinates as estimated by each DWC coincide at the level of 1.5 mm (which corresponds to a rough estimation of the DWC intrinsic resolution). This will be referred to as the ``DWC selection'' in the following. 

The small muon component of the beam is further suppressed by requiring that the signal in the muon counter scintillator is compatible with its pedestal. 

The selection of positrons is completed by the request of a large signal in the threshold Cherenkov counters. Several combinations of selection criteria were tested, including those making use of the PS,  and the final one was chosen by balancing  the efficiency and purity of the positron selection. The chosen selection criteria require a signal higher than $3\sigma_{\mathrm{ped}}$ in one Cherenkov counter. 

\begin{figure}[htb]
\begin{center}
\includegraphics[width=0.6\textwidth]{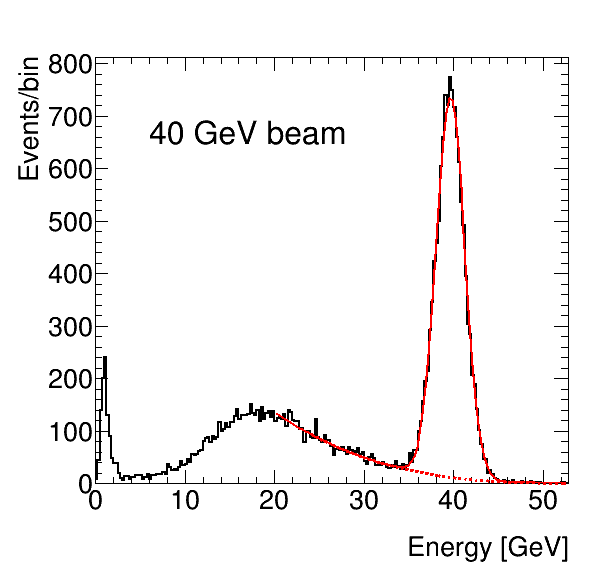}
\caption{Energy distribution in the calorimeter (sum of Cherenkov and scintillation signals) after the DWC selection. The result of a fit with a third degree polynomial is shown by a dashed red line, while that of a fit with a Gaussian plus a third degree polynomial is shown as a solid red line.}
\label{fig:purity_40GeV}
\end{center}
\end{figure}

The beam purity was estimated at all energies by fitting the energy distribution in the calorimeter after energy equalisation (discussed in Section~\ref{sec:calo_equalise}) with a third degree polynomial for the non-electron component (mainly residual hadrons) and a Gaussian for the electrons. The beam purity is defined as the ratio of the integral of the Gaussian peak and the total number of selected events. The resulting fit is shown in Figure~\ref{fig:purity_40GeV} for $\Ebeam = 40 \ \mathrm{GeV}$ as an example. 

The beam purity estimated after applying only the DWC selection is reported in Table~\ref{tab:beam_purity} for a subset of the energies considered. The beam purity was found to vary little between 40 and 100 GeV.

\begin{table}[h!]
\centering
\caption{Fraction of positrons in the beam as a function of the beam energy \Ebeam. The purity was estimated after applying the DWC selection described in the text.}
\vspace{0.5cm}
\begin{tabular}{c|c}
\hline
\hline
\Ebeam [GeV] & Positron Purity \\ \hline
\hline
10 & 65\% \\ \hline
20 & 57\% \\ \hline
40 & 50\% \\ \hline
100 & 46\% \\ \hline
120 & 7.5\% \\ \hline\hline
\end{tabular}
\label{tab:beam_purity}
\end{table}

The purity was estimated again with the same method after the full positron selection described in the text. It was determined to be above 95\% at all energies considered. Possible systematics associated with the residual contamination from non-positron components on the measurements of Section~\ref{sec:results} were evaluated to be negligible.

\section{Detector calibration}
\label{sec:calibration}

The calibration of the prototype was performed in several steps. First, the gain of all SiPMs in $\M{0}$ was equalised, and the conversion factor between ADC counts and \phe was derived, by making use of the SiPM multiphoton spectrum, as discussed in Section~\ref{sec:multiphoton}. Then, the response of all modules $\M{0}$-$\M{8}$ was roughly equalised by making use of a positron beam with an energy of 20 GeV (Section~\ref{sec:calo_equalise}). After equalisation, the overall calorimeter energy scale was set by looking at the response of the whole calorimeter prototype to beams of positrons (Section~\ref{sec:calo_calibrate}). On top of that, a set of tower-level energy corrections was derived at analysis level, as described in Section~\ref{sec:offline_calibration}.  The contribution of the electronic noise to the energy measurement is discussed in Section~\ref{sec:pedestals}.

\subsection{SiPM equalisation using the multiphoton spectrum}
\label{sec:multiphoton}

A first tuning of the amplifier settings to equalise the SiPM response was done by using an ultra-fast LED emitting at 420 nm in the labs before the test beam period. The same voltage of $+7\ \mathrm{V}$ over breakdown was applied to all SiPMs. The setting is not typical for a SiPM, but it guarantees a Photon Detection Efficiency (PDE) stable under small temperature variations, and a charge multiplication factor of about $0.5 \times 10^{6}$ for each detected photon. 

\begin{figure}[ht]
\begin{center}
    \includegraphics[width=0.70\textwidth]{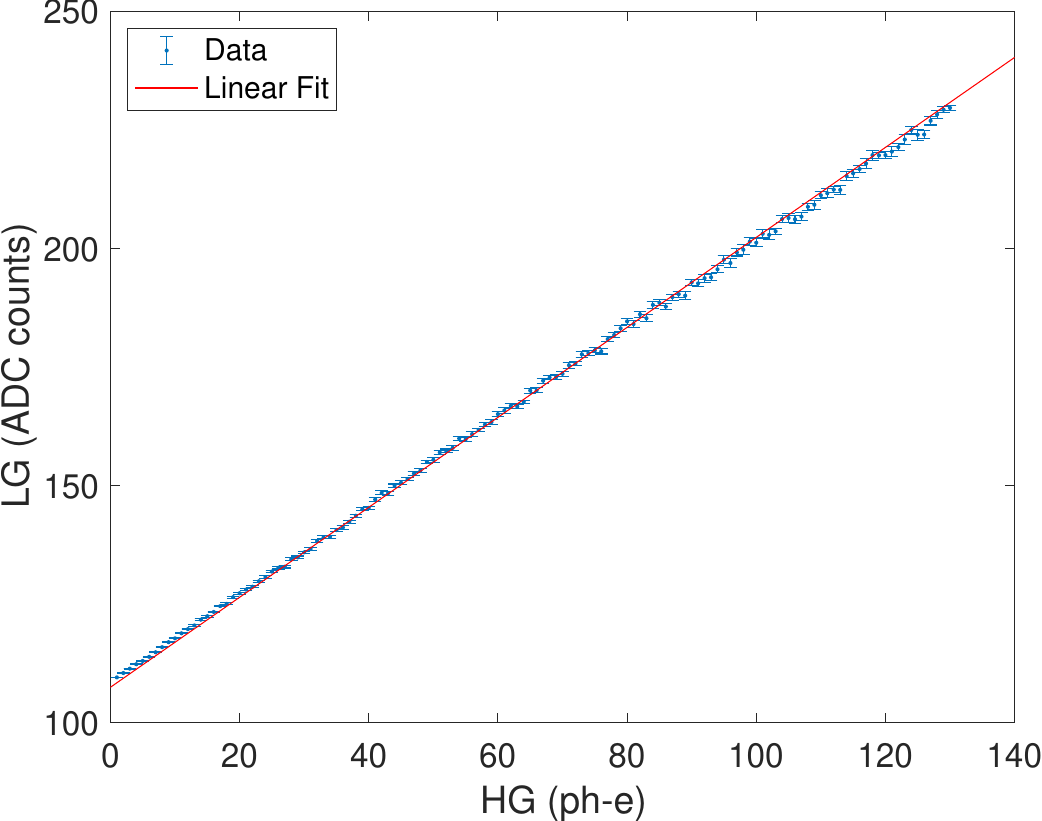}
\caption{Response in ADC counts for the LG against that in \phe for the HG for one example SiPM.}
\label{fig:calibration}
\end{center}
\end{figure}

The multiphoton spectrum recorded with HG is the starting point of the in-situ SiPM equalisation procedure: the procedure is similar to that discussed in Ref.~\cite{Ampilogov:2023zxb}, where full details can be found. For each SiPM:

\begin{itemize}
\item The peaks of the multiphoton spectrum in HG were fitted with Gaussian distributions, and from the peak-to-peak distance a conversion factor from ADC to \phe was determined. A typical conversion factor was about 0.05 \phe/ADC count with uncertainties of the order of 0.1\%.   
\item The pedestal of the HG was determined from a fit to the pedestal peak of the multiphoton spectrum. The pedestal of the LG was extracted in a similar way after selecting events which are in the pedestal peak of the HG, to avoid possible biases.  
\item The signal in the SiPM corresponding to 40-GeV positron events was recorded for both the LG and the HG. The LG signal was plotted against that of the HG, as illustrated in \Figure~\ref{fig:calibration} for one specific SiPM. A linear fit (with the pedestals fixed to those extracted independently) determined a conversion factor from HG to LG\footnote{The fit to \Figure~\ref{fig:calibration} can also be done leaving the pedestals of LG floating to be determined by the fit. This was tried, and the results are consistent with those extracted independently.}. 
\end{itemize}

The parameters were extracted multiple times during the test beam, and the findings about the stability of the SiPM calibrations of  Ref.~\cite{Ampilogov:2023zxb} were confirmed. Given the stability of 
 the calibration parameters over time~\cite{instruments6040059}, a single set of calibration constants for each SiPM was used for the whole data taking period. For a given SiPM, the signal used in the following for the data analysis is that from the HG, unless found to be saturated, in which case the LG signal is used. 

 It is well known~\cite{Stoykov:2007hs} that SiPMs yield a non-linear response to the incoming light when the number of photons in the pulse is a significant fraction of the number of cells available for the SiPM. In this case, a standard procedure is to correct the SiPM response using the following formula: 

 \begin{align*} 
 N_{\mathrm{fired}} = N_{\mathrm{cells}} \times \left( 1-e^{-\frac{N_{\mathrm{photons}} \times \mathrm{PDE}}{N_{\mathrm{cells}}}} \right)
 \end{align*}

 Here $N_{\mathrm{cells}}$ is the number of cells available for the SiPM (7284 for the S14160-1315), $N_{\mathrm{fired}}$ is the recorded signal in \phe, $N_{\mathrm{photons}}$ is the number of photons that hit the cathode. By inverting this relation, the SiPM response can be corrected to account for non-linearities. This procedure was implemented to correct for the SiPM response. For example, this correction was at the level of 5\% when a signal of 2 GeV was measured in a single scintillating fibre. 

\subsection{Calorimeter response equalisation}
\label{sec:calo_equalise}

Next, the response of the $\M{0}-\M{8}$ modules was equalised using a set of runs using a beam of positrons with a momentum of 20 GeV. In each run, the module was positioned on beam so that the shower barycentre matched the geometrical centre of each module. A well-centred, high-purity positron beam can be obtained by applying the positron selection of Section~\ref{sec:part_selection}. The equalisation procedure assumed an equal tower response to positrons in such conditions. The equalisation was obtained by setting the response of all modules equal to that of $\M{0}$. In other words, if we define the average response in \phe of $\M{0}$ when hit in its centre by a beam of 20-GeV positrons as $P^{S,C}_{\M{0}} \left(20\ \mathrm{GeV}\right)$ (where the letter $S$ or $C$ represents scintillation or Cherenkov), then the average response (in \phe) of the $i$-th module $\M{i}$ is obtained by aiming the beam at its own centre, measuring its average response in ADC $A^{S,C}_{\M{i}}$, and computing a constant $a_i^{S,C}$ so that 

\begin{align*}
    P^{S,C}_{\M{i}} \left(20\ \mathrm{GeV}\right) = a_i^{S,C} \times A^{S,C}_{\M{i}}\ \left(20\ \mathrm{GeV}\right) = P^{S,C}_{\M{0}} \left(20\ \mathrm{GeV}\right)\quad (i\in [1,8]),
\end{align*}

The containment of a single module to a 20-GeV positron beam is estimated to be $\epsilon_{\mathrm{module}} = 72\%$ with the help of the Geant4 test beam simulation. Using this information and the response in \phe, one can compute the light yield per unit of energy deposited in the prototype: it is about 290 (70) \phe/GeV for the scintillation (Cherenkov) readout.

\subsection{Calorimeter calibration}
\label{sec:calo_calibrate}

The overall calorimeter electromagnetic energy scale was set by rescaling the sum of the average responses on \phe of the modules $\M{0}-\M{8}$ by a single pair of common constants $\delta_{S}$ for the scintillation signal and $\delta_{C}$ for the Cherenkov signal, so that the total energy measured in the calorimeter corresponded to the beam energy separately for the scintillation and Cherenkov signal\footnote{The simulation predicts a shower containment of the full prototype of $\epsilon = 94\%$ at $\Ebeam = 20 \ \mathrm{GeV}$. This number was found to be nearly independent of $\Ebeam$. This lateral leakage was therefore de-facto reabsorbed in the determination of $\delta_{S,C}$.}. The constants $\delta_{S}$ and $\delta_{C}$ are determined as 

\begin{align*}
    \delta_{S,C} = \frac{20\ \mathrm{GeV}}{\langle\sum_{i = 0}^{8} P^{S,C}_{\M{i}}\rangle},
\end{align*}

\noindent where the average is computed over all selected positrons from a 20-GeV run with the beam pointing to the geometric centre of $\M{0}$.

\subsection{Offline analysis calibration}
\label{sec:offline_calibration}

After the data taking was completed, it was noted that there was a small offset, at the level of about 5\%, between the two readout responses (Cherenkov and scintillation). A possible explanation may stem from the fact that the shower containment of a single module as seen by the scintillation or Cherenkov readout is different (the shower is wider for the Cherenkov component, leading to a larger lateral leakage~\cite{Ampilogov:2023zxb}).  

A final calibration step was therefore performed during the analysis phase, independently for the Cherenkov and scintillation readouts. The total energy in a given event is defined as 

\begin{align*} 
E_{S,C} = \delta_{S,C} \sum_0^8 \beta^{S,C}_i P^{S,C}_{\M{i}} + \beta_{\mathrm{Ps}} S_{\mathrm{Ps}},
\end{align*}

\noindent where $S_{\mathrm{Ps}}$ is the signal recorded for the preshower scintillator: it is included in the calculation to account for fluctuations in the electron energy loss in the preshower material. 

The coefficients $\beta^{S,C}_i$ are computed analytically by minimising the RMS of the residuals with respect to the nominal beam energy. The computation was done by using runs of positrons with an energy of 20 GeV oriented so that the shower barycentre coincides with the nominal centre of $\M{0}$.

The values of the 18 $\beta^{S,C}_i$ are between 0.85 and 1.15. The contribution of the energy loss in the preshower detector is estimated to be $30\ \mathrm{MeV/MIP}$.

\begin{figure}[ht]
\begin{center}
    \includegraphics[width=0.60\textwidth]{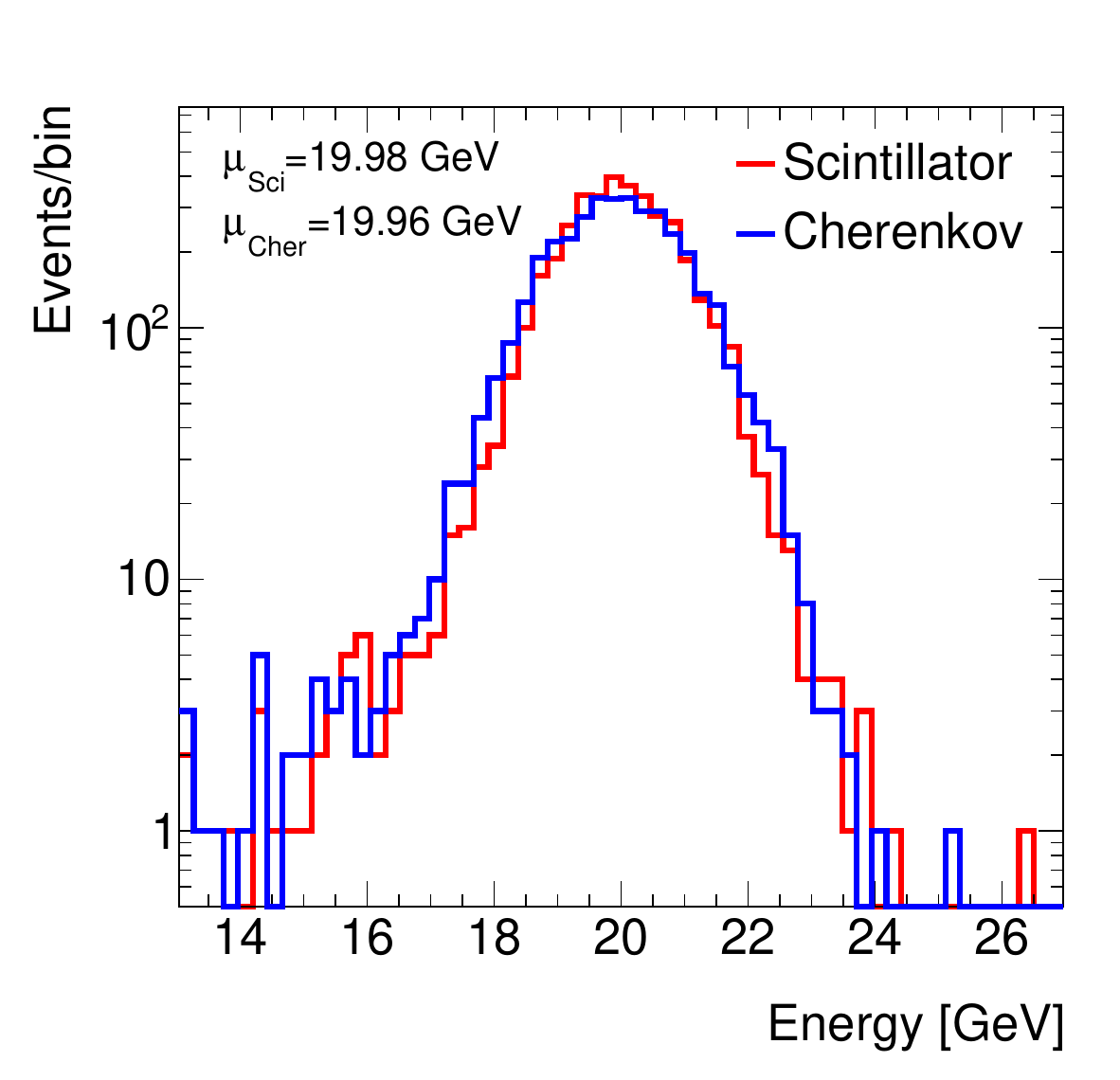}
\caption{Energy measurement of the calorimeter using a 20 GeV positron beam after the full calibration procedure is applied. The results are shown separately for the Cherenkov  (in blue) and scintillation (in red) channels.}
\label{fig:energy_response}
\end{center}
\end{figure}

The energy distribution as measured by the calorimeter in response to a 20 GeV positron beam is shown in Figure~\ref{fig:energy_response}.

\subsection{Noise determination}
\label{sec:pedestals}

The electronic noise contribution from the PMTs reading the modules \M{1}-\M{8} was estimated simply by making use of the pedestal triggers. The RMS of the pedestal distribution was found to be 110 (150) MeV for the sum of the scintillation (Cherenkov) PMT channels. The combined noise contribution was determined as the RMS of the distribution of the sum of all PMTs of the scintillation and Cherenkov channels divided by two, and it was found to be about 120 MeV.

The estimate of the SiPMs contribution to the noise is less straightforward: because of the self-triggering system of the FERS, there is essentially no data from the FERS associated with pedestal triggers. However, the multiphoton spectrum for HG can still be measured in physics events, allowing the determination of the individual channel pedestal RMS. The noise was determined to be about 2 MeV for each HG Cherenkov channel and 0.6 MeV for each HG scintillation channel. By selecting events contributing to the pedestal of the HG, the LG pedestal width can also be measured, and it was found to be about 50 MeV for each Cherenkov channel, and about  12 MeV for the scintillation channels. The apparent difference in the value stems from the different calibration factors applied to the LG and HG channels. 

By summing the HG signals that measure a signal compatible with pedestal, it is possible to check the dependency of the total SiPM channel noise on the number of SiPM channels considered. It was found that the noise scales nearly linearly when summing SiPM channels on the same FERS board~\cite{Stoykov:2007hs}. There is therefore evidence that the noise is highly correlated between channels read out by the same FERS board. By doing similar tests, the level of correlation between FERS boards can be checked: the noise was found to be only loosely correlated for channels located on different FERS. The exact level of channel-by-channel correlation depends on the FERS board considered.

\section{Results}
\label{sec:results}
Following calibration, the data were analysed, and, similarly to Ref.~\cite{Ampilogov:2023zxb}, the performance of the prototype was characterised in terms of linearity and resolution of the energy response. For the latter in particular, the much improved positron beam purity at all energies, together with a closer positioning of the preshower detector to the calorimeter led to a complete assessment of the calorimeter resolution to positrons with beam energies from 10 to 120 GeV.    

\subsection{Positron energy measurement}

\begin{figure}[htb]
\begin{center}
        \includegraphics[width=0.47\textwidth]{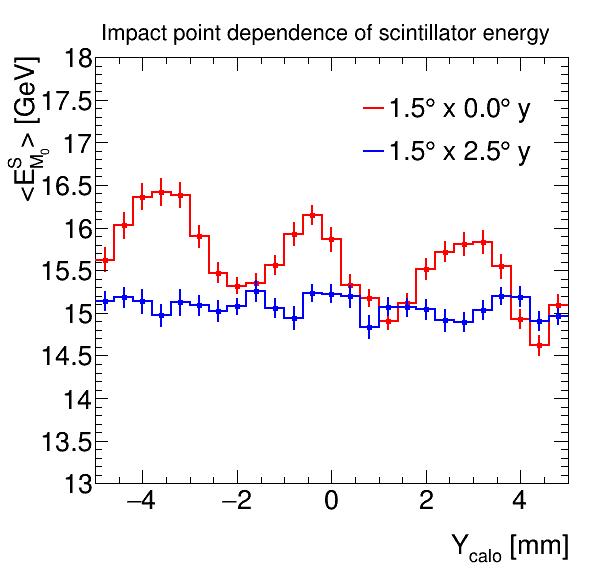}
\caption{Distribution of the average scintillator energy deposition in the central module in bins of the y coordinate of the beam impact point on the calorimeter for two different impact angles in the $y-z$ plane.}
\label{fig:angledep}
\end{center}
\end{figure}

As predicted in Ref.~\cite{Ampilogov:2023zxb}, the positioning of the calorimeter with a tilt angle both in the $x-z$ and in the $y-z$ plane led to a calorimeter response independent of the impact point of the particle on the calorimeter's front face. This is shown in Figure~\ref{fig:angledep}, where 
the average scintillator energy deposition in \M{0} 
is shown as a function of the vertical coordinate of the beam impact point for two different impact angles on the calorimeter. A clear modulation is seen when the beam is aligned with the fibres in the y direction, while no modulation is observed when the calorimeter is inclined by 2.5 degrees. 

The combined dual-readout response of the calorimeter $E$ was computed as the arithmetic average of the Cherenkov and scintillating channels, $E = (E_S+E_C)/2$. 

\begin{figure}[htb]
\begin{center}
\subfigure[]{
        \includegraphics[width=0.47\textwidth]{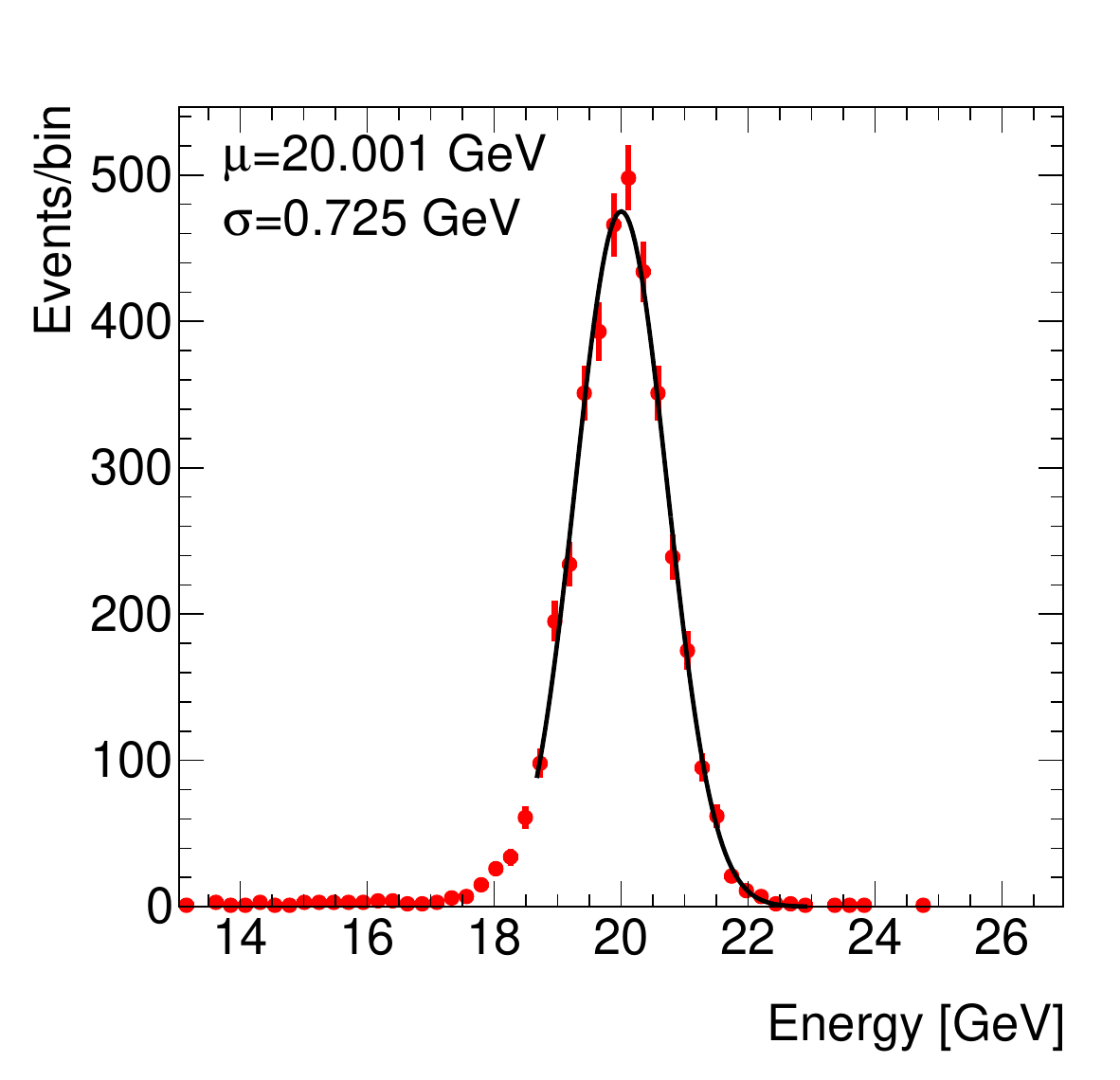}
        } \hfill
\subfigure[]{
        \includegraphics[width=0.47\textwidth]{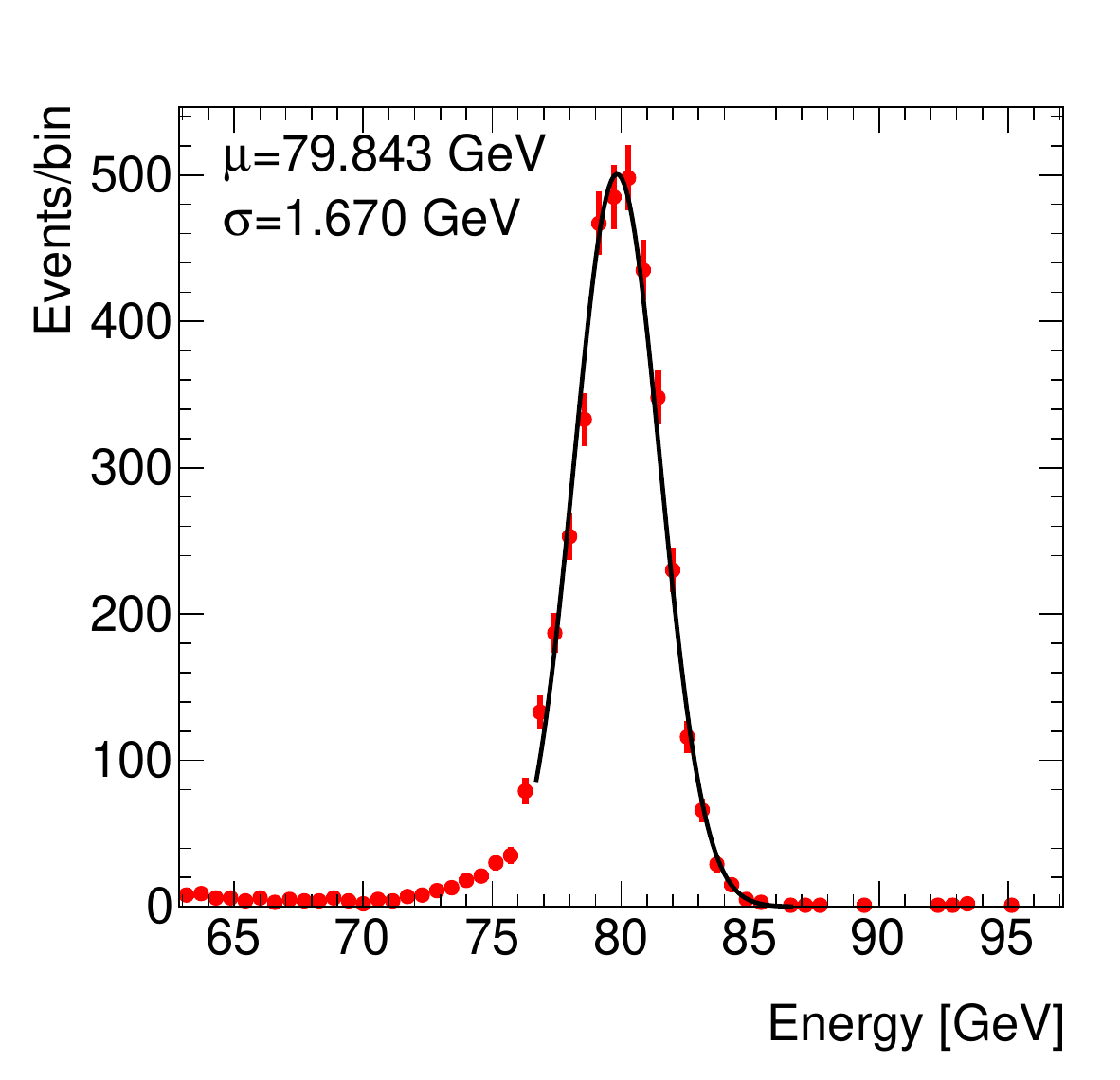}
        }
\caption{Distribution of the combined energy response for positron beam energies of (a) 20 GeV and (b) 80 GeV.}
\label{fig:example_response}
\end{center}
\end{figure}

As an example, the combined energy response to positron beams of 20 and 80 GeV is shown in Figure~\ref{fig:example_response}.   

The distributions are centred at the nominal beam energies. The distributions are very close to Gaussians in their core. Similar histograms were produced for the individual scintillation and Cherenkov channels, and for all the available beam energies. In each case, the histograms were fit with a Gaussian between $m - \alpha_{\mathrm{low}}\times r$ and $m + \alpha_{\mathrm{high}} \times r$, where $m$ represents the mean value of the distribution and $r$ its RMS, $\alpha_{\mathrm{low}} = 1.8$ and $\alpha_{\mathrm{high}} = 4$. We name the mean of the fitted Gaussian as $\Emeas$ and its width as $\sigma_t$. The fit interval was chosen to exclude the low-energy tail (due to a non-complete rejection of particles other than positrons in the beam) present in the distributions. Different values for $\alpha_{\mathrm{low}} \in [1,2]$ and $\alpha_{\mathrm{high}} \in [1,5]$ were tried, and the specific choice does not significantly affect the conclusions on the performance of the response measurement.

\begin{figure}[htb]
\begin{center}
\subfigure[]{
	\includegraphics[width=0.48\textwidth]{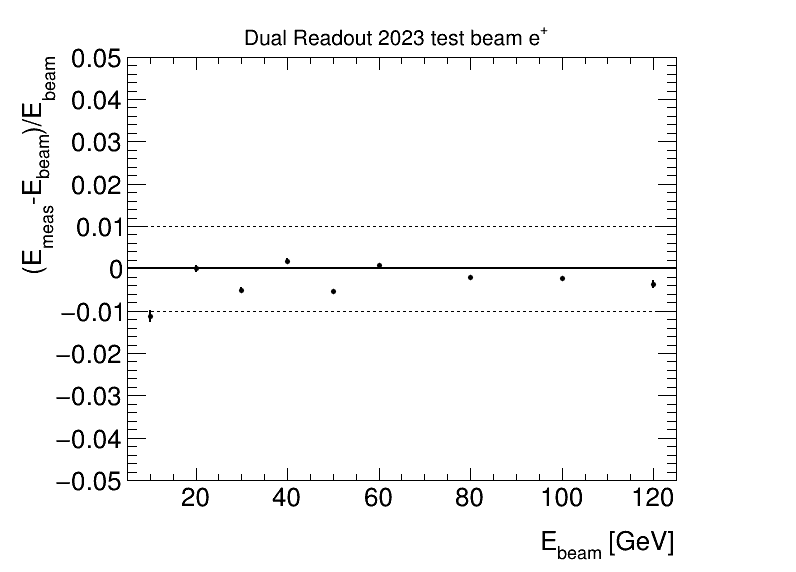}
 } \hfill
 \subfigure[]{
	\includegraphics[width=0.48\textwidth]{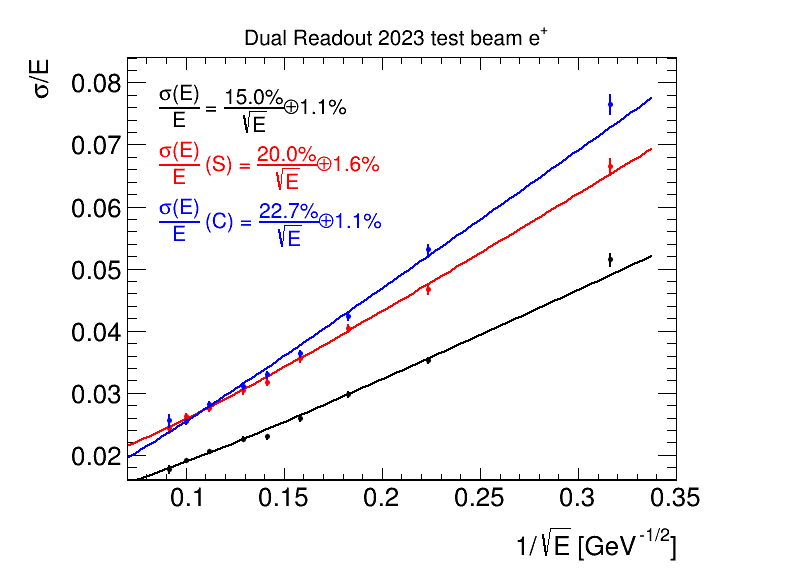}
 }
\caption{(a) Linearity and (b) noise-subtracted resolution of the calorimeter response as a function of the beam energy. For (b), the independent resolutions of the Cherenkov (in blue) and scintillation (in red) channels are shown, together with that of the combined response (in black).  
}
\label{fig:calolinres}
\end{center}
\end{figure}

 The linearity was studied defining the fractional difference between $\Emeas$ and the beam energy, $(\Emeas - \Ebeam)/\Ebeam$. It is shown as a function of \Ebeam in Figure~\ref{fig:calolinres} (a). The bias was found to be 1\% or less.

One of the key improvements of this paper with respect to Ref.~\cite{Ampilogov:2023zxb} is the assessment of the calorimeter resolution with real data. In Ref.~\cite{Ampilogov:2023zxb}, the non-optimal experimental conditions described before forced us to assess the calorimeter energy resolution to positrons at energies above 30 GeV by making use of the test beam software simulation, after having verified that the results were compatible with real data at low energies.

To evaluate the resolution of the energy response, the values of $\sigma_t$ were corrected for the effect of the electronic noise discussed in Section~\ref{sec:pedestals}. A model for the expected noise on the individual scintillation and Cherenkov channels and for the combined response was built by taking into account the PMT noise, the individual SiPM noise for HG and LG, the number of HG and LG channels used at each energy, the correlation between channels read out by the same FERS. The values of the noise for the combined response span between $\sigma_{\mathrm{noise}} = 120 \ \mathrm{MeV}$ at $\Ebeam = 10\ \mathrm{GeV}$ and $\sigma_{\mathrm{noise}} = 280 \ \mathrm{MeV}$ at $\Ebeam = 120\ \mathrm{GeV}$. 

We define the resolution of the energy measurement as the ratio of $\sigma = \sqrt{\sigma_t^2 - \sigma^2_{\mathrm{noise}}}$ to $\Emeas$\footnote{The noise contribution is subtracted from the data (rather than fitted as an additional term proportional to $1/E$) because it depends on $E_{\mathrm{beam}}$.}  It is shown in Figure~\ref{fig:calolinres} (b). The Cherenkov, scintillation, and combined resolutions as a function of the beam energy are fit with a function of the type

\begin{align*} 
\frac{\sigma}{\Emeas} = \frac{a}{\sqrt{\Emeas}}\oplus c
\end{align*}

\noindent where $\Emeas$ is expressed in GeV. The fit to the combined response curve yielded an estimate of the value of the stochastic term $a = 15.0\%$ and of the constant term  $c = 1.1\%$.

\begin{figure}[htb]
\begin{center}

\includegraphics[width=0.6\textwidth]{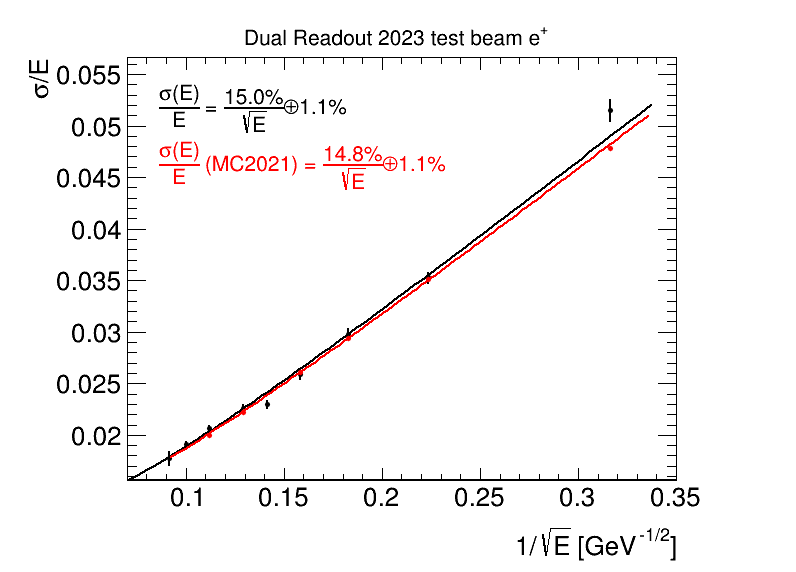}
\caption{Distribution of the energy measured after all corrections for a 20-GeV 
positron beam for (a) data and (b) the simulation.}
\label{fig:resol_comparewithmc}
\end{center}
\end{figure}

The measured resolution curve can be compared with the test beam software simulation 
obtained in Ref.~\cite{Ampilogov:2023zxb}, after the incorporation into the simulation of a spread on the beam energy of 1\% as suggested by the CERN beam experts. The result is shown in Figure~\ref{fig:resol_comparewithmc}, which shows an excellent agreement between data and simulation, except for the 10~GeV point, which may indicate a residual energy-independent term in the resolution. 

\section{Conclusions}
\label{sec:conclusions}
A dual-readout sampling calorimeter prototype using brass capillary tubes as absorber and optical fibres as active medium was tested using beams of particles at  the H8 beam line at CERN. The dual readout was realised by making use of two different types of fibres: doped scintillating Saint-Gobain BCF-10 fibres, and clear ``Cherenkov'' Mitsubishi SK40 fibres. The prototype (with a total size of about $10\times10\times 100\ \mathrm{cm^3}$) was composed of nine modules. For the central module, the individual fibres were read out by means of Hamamatsu S14160-1315 PS SiPMs, while for the surrounding eight modules the two sets of fibres were bundled together and read out by Hamamatsu R8900 PMTs. 

The detector was calibrated by making use of the SiPM multiphoton spectrum and of beams of positrons. Then, the detector response was studied using beams of positrons with energies between 10 and 120~GeV. Thanks to the excellent beam purity, at all beam energies, it was possible to assess the detector response in terms of linearity and resolution of the energy measurement. The linearity was found to be within 1\%. The energy resolution was found to be 

\begin{align*}
    \frac{\sigma}{E} = \frac{15.0\%}{\sqrt{E\ [\mathrm{GeV}]}} \oplus 1.1\%,
\end{align*}

\noindent in agreement with what was estimated with a Geant4 simulation of the detector validated at previous test-beams for ideal test-beam conditions, after taking into account the beam momentum spread.  

The results on the electromagnetic performance of the dual-readout sampling calorimeter described in this paper confirm that the capillary tube mechanical solution in conjunction with a SiPM based readout is a viable solution for future developments, and pave the way to the use of this technology in a prototype big enough to largely contain the hadronic shower. The work is ongoing under the HiDRa project~\cite{Santoro:2024aei}.

\acknowledgments{The authors would like to thank the CERN SPS team for their assistance and for providing excellent beam quality during our test experiments. This project has received funding from the European Union’s Horizon 2020 Research and Innovation programme under Grant Agreements No 101004761 and 101057511.}

\bibliography{DRtubes}

\providecommand{\href}[2]{#2}\begingroup\raggedright\begin{thebibliography}{10}

\bibitem{Lee:2017xss}
S.~Lee, M.~Livan and R.~Wigmans, \emph{{Dual-Readout Calorimetry}},
  \href{https://doi.org/10.1103/RevModPhys.90.025002}{\emph{Rev. Mod. Phys.}
  {\bfseries 90} (2018) 025002}
  [\href{https://arxiv.org/abs/1712.05494}{{\ttfamily 1712.05494}}].

\bibitem{FCC:2018evy}
{\scshape FCC} collaboration, \emph{{FCC-ee: The Lepton Collider}: {Future
  Circular Collider Conceptual Design Report Volume 2}},
  \href{https://doi.org/10.1140/epjst/e2019-900045-4}{\emph{Eur. Phys. J. ST}
  {\bfseries 228} (2019) 261}.

\bibitem{CEPCStudyGroup:2018ghi}
{\scshape CEPC Study Group} collaboration, \emph{{CEPC Conceptual Design
  Report: Volume 2 - Physics \& Detector}},
  \href{https://arxiv.org/abs/1811.10545}{{\ttfamily 1811.10545}}.

\bibitem{Akchurin:2005eu}
N.~Akchurin et~al., \emph{{Electron detection with a dual-readout
  calorimeter}}, \href{https://doi.org/10.1016/j.nima.2004.06.178}{\emph{Nucl.
  Instrum. Meth. A} {\bfseries 536} (2005) 29}.

\bibitem{Akchurin:2005an}
N.~Akchurin et~al., \emph{{Hadron and jet detection with a dual-readout
  calorimeter}}, \href{https://doi.org/10.1016/j.nima.2004.07.285}{\emph{Nucl.
  Instrum. Meth. A} {\bfseries 537} (2005) 537}.

\bibitem{Akchurin:2005rs}
N.~Akchurin et~al., \emph{{Comparison of high-energy electromagnetic shower
  profiles measured with scintillation and Cherenkov light}},
  \href{https://doi.org/10.1016/j.nima.2005.03.169}{\emph{Nucl. Instrum. Meth.
  A} {\bfseries 548} (2005) 336}.

\bibitem{Akchurin:2013yaa}
N.~Akchurin et~al., \emph{{Dual-readout Calorimetry}},
  \href{https://arxiv.org/abs/1307.5538}{{\ttfamily 1307.5538}}.

\bibitem{Lee:2017shn}
S.~Lee et~al., \emph{{Hadron detection with a dual-readout fiber calorimeter}},
  \href{https://doi.org/10.1016/j.nima.2017.05.025}{\emph{Nucl. Instrum. Meth.
  A} {\bfseries 866} (2017) 76}
  [\href{https://arxiv.org/abs/1703.09120}{{\ttfamily 1703.09120}}].

\bibitem{Akchurin:2014aoa}
N.~Akchurin et~al., \emph{{The electromagnetic performance of the RD52 fiber
  calorimeter}}, \href{https://doi.org/10.1016/j.nima.2013.09.033}{\emph{Nucl.
  Instrum. Meth. A} {\bfseries 735} (2014) 130}.

\bibitem{Antonello:2018sna}
M.~Antonello et~al., \emph{{Tests of a dual-readout fiber calorimeter with SiPM
  light sensors}},
  \href{https://doi.org/10.1016/j.nima.2018.05.016}{\emph{Nucl. Instrum. Meth.
  A} {\bfseries 899} (2018) 52}
  [\href{https://arxiv.org/abs/1805.03251}{{\ttfamily 1805.03251}}].

\bibitem{IDEAStudyGroup:2025gbt}
{\scshape IDEA Study Group} collaboration, \emph{{The IDEA detector concept for
  FCC-ee}},  \href{https://arxiv.org/abs/2502.21223}{{\ttfamily 2502.21223}}.

\bibitem{lorenzoPhD}
L.Pezzotti, \emph{Particle Detectors R\&D: Dual-Readout Calorimetry for Future
  Colliders and MicroMegas Chambers for the ATLAS New Small Wheel Upgrade},
  Ph.D. thesis, {Universit\'a degli Studi di Pavia}, 2021.

\bibitem{Pezzotti:2022ndj}
I.~Pezzotti et~al., \emph{{Dual-Readout Calorimetry for Future Experiments
  Probing Fundamental Physics}},
  \href{https://arxiv.org/abs/2203.04312}{{\ttfamily 2203.04312}}.

\bibitem{Lucchini:2020bac}
M.T.~Lucchini et~al., \emph{{New perspectives on segmented crystal calorimeters
  for future colliders}},
  \href{https://doi.org/10.1088/1748-0221/15/11/P11005}{\emph{JINST} {\bfseries
  15} (2020) P11005} [\href{https://arxiv.org/abs/2008.00338}{{\ttfamily
  2008.00338}}].

\bibitem{Lucchini:2022vss}
M.T.~Lucchini, L.~Pezzotti, G.~Polesello and C.G.~Tully, \emph{{Particle flow
  with a hybrid segmented crystal and fiber dual-readout calorimeter}},
  \href{https://doi.org/10.1088/1748-0221/17/06/P06008}{\emph{JINST} {\bfseries
  17} (2022) P06008} [\href{https://arxiv.org/abs/2202.01474}{{\ttfamily
  2202.01474}}].

\bibitem{Karadzhinova-Ferrer:2022paf}
A.~Karadzhinova-Ferrer et~al., \emph{{Novel prototype tower structure for the
  dual-readout fiber calorimeter}},
  \href{https://doi.org/10.1088/1748-0221/17/09/T09007}{\emph{JINST} {\bfseries
  17} (2022) T09007}.

\bibitem{Ampilogov:2023zxb}
N.~Ampilogov et~al., \emph{{Exposing a fibre-based dual-readout calorimeter to
  a positron beam}},
  \href{https://doi.org/10.1088/1748-0221/18/09/P09021}{\emph{JINST} {\bfseries
  18} (2023) P09021} [\href{https://arxiv.org/abs/2305.09649}{{\ttfamily
  2305.09649}}].

\bibitem{6551127}
D.L.~Chichester, S.M.~Watson and J.T.~Johnson, \emph{Comparison of bcf-10,
  bcf-12, and bcf-20 scintillating fibers for use in a 1-dimensional linear
  sensor},  in \emph{2012 IEEE Nuclear Science Symposium and Medical Imaging
  Conference Record (NSS/MIC)}, pp.~365--370, 2012,
  \href{https://doi.org/10.1109/NSSMIC.2012.6551127}{DOI}.

\bibitem{SK40}
{Mitsubishi ESKA SK40}.
  \url{https://www.pofeska.com/pofeskae/download/pdf/f/SK40.pdf}, accessed on
  14 March 2023.

\bibitem{R8900}
{HAMAMATSU R8900}.
  \url{https://www.hamamatsu.com/content/dam/hamamatsu-photonics/sites/documents/99_SALES_LIBRARY/etd/R8900(U)-00-C12_TPMH1299E.pdf},
  accessed on 14 March 2023.

\bibitem{S14160}
{HAMAMATSU S14160-1315PS}.
  \url{https://www.hamamatsu.com/jp/en/product/optical-sensors/mppc/mppc_mppc-array/S14160-1315PS.html},
  accessed on 16 March 2023.

\bibitem{FERS:CAEN}
{FERS A5202, CAEN S.p.A., FERS A5202}.
  \url{https://www.caen.it/products/a5202/}, accessed on 18 January 2023.

\bibitem{CITIROC}
Citiroc-1A. \url{https://www.weeroc.com/products/sipm-read-out/citiroc-1a},
  accessed on 18 January 2023.

\bibitem{Dannheim:2013iea}
{\scshape Linear Collider CERN Detector} collaboration, \emph{{Particle
  Identification with Cherenkov detectors in the 2011 CALICE Tungsten Analog
  Hadronic Calorimeter Test Beam at the CERN SPS}}, {\emph{LCD-Note-2013-006,
  AIDA-NOTE-2015-012} (2013) }.

\bibitem{instruments6040059}
R.~Santoro, \emph{Sipms for dual-readout calorimetry},
  \href{https://doi.org/10.3390/instruments6040059}{\emph{Instruments}
  {\bfseries 6} (2022) }.

\bibitem{Stoykov:2007hs}
A.~Stoykov, Y.~Musienko, A.~Kuznetsov, S.~Reucroft and J.~Swain, \emph{{On the
  limited amplitude resolution of multipixel Geiger-mode APDs}},
  \href{https://doi.org/10.1088/1748-0221/2/06/P06005}{\emph{JINST} {\bfseries
  2} (2007) P06005} [\href{https://arxiv.org/abs/0706.0746}{{\ttfamily
  0706.0746}}].

\bibitem{Santoro:2024aei}
{\scshape IDEA Dual-Readout group} collaboration, \emph{{HiDRa -
  High-resolution Calorimeter for e+e-}},
  \href{https://doi.org/10.22323/1.476.1068}{\emph{PoS} {\bfseries ICHEP2024}
  (2025) 1068}.

\end{thebibliography}\endgroup
\bibliographystyle{JHEP}

\end{document}